\documentclass[11pt,tightenlines,nofootinbib,a4paper]{revtex4}
\usepackage{amsmath,amscd,amssymb}
\usepackage{epsfig,color,eso-pic}
\usepackage{graphics,mathtools}
\usepackage{diagbox}

\newcommand{\imu}{{\rm i}}

\newcommand{\zr}[1]{\mbox{\hspace*{#1em}}}
\newcommand{\ID}{\mbox{{\sf 1}\zr{-0.16}\rule{0.04em}{1.55ex}\zr{0.1}}}

\begin{document}

\title{Quantum energies of solitons with different topological charges}

\author{N. Graham$^{a)}$, H. Weigel$^{b)}$}

\affiliation{
$^{a)}$Department of Physics, Middlebury College
Middlebury, VT 05753, USA\\
$^{b)}$Institute for Theoretical Physics, Physics Department,
Stellenbosch University, Matieland 7602, South Africa}

\begin{abstract}
The vacuum polarization energy is the leading quantum correction to the 
classical energy of a soliton. We study this energy for two-component solitons
in one space dimension as a function of the soliton's topological 
charge. We find that both the classical and the vacuum polarization
energies are linear functions of the topological charge with a 
small offset. Because the combination of the classical and quantum 
offsets determines the binding energies, either all higher charge solitons 
are energetically bound or they are all unbound, depending on model
parameters. This linearity persists even when the field configurations
are very different from those of isolated solitons, and would not be
apparent from an analysis of their bound state spectra alone.
\end{abstract}

\maketitle

\section{Introduction}

Solitons (or solitary waves) are solutions to non-linear wave equations
with a localized energy density and which thus have a particle interpretation.
Typically solitons fall into topological sectors characterized by integer
charges with an infinite energy barrier between 
different sectors. The classical energy of solitons grows 
with these topological charges: often models have a lower, so-called 
Bogomolny-Prasad-Sommerfield (BPS) bound that is linear in the (absolute value
of the) topological charge~\cite{Bogomolny:1975de,Prasad:1975kr}. Field 
configurations with higher topological charges fall into the same topological
sector as equally many widely separated solitons with unit
charge. The stability of the former is then decided by the energy balance 
when comparing configurations with (almost) equal classical energies.

These classical energies have quantum corrections. On the absolute scale they 
should be small for a consistent model, but they may be important for the 
energy balance. Because these corrections emerge from a quantum field theory 
calculation, only renormalizable models can be considered.  Unfortunately, 
renormalizable models with exact static solitons of different topological
charges are very rare. Recently we have computed the one-loop quantum
corrections to BPS vortices in scalar electrodynamics \cite{Graham:2022adn}.
In this model the numerics are very intricate and therefore only
winding number up to four was considered. Within that range, the
quantum correction was essentially linear in the topological charge,
with a small offset. There are many renormalizable (at least to one
loop) soliton models in one space dimension, but most of them allow
only a unit topological charge. The sine-Gordon model is an exception,
but its static classical multi-charge solutions merely consist
of single charge configurations that do not interact.  In a recently 
proposed \cite{Halcrow:2023eph} two-field generalization of the sine-Gordon 
model, however, a second field produces the necessary attraction between the 
single charge configurations so that multi-charge solutions can emerge as a 
single bound lump. We will investigate this model to compute the leading 
quantum corrections to the energies of the configurations of the higher 
charge solutions and determine their binding energies.

Similar extensions of the $\phi^4$ and $\phi^6$ kink models have also recently 
been proposed \cite{Halcrow:2022kfo}. However, they have a limited range of 
integer topological charges and we will thus not consider them here.

\section{The model and its soliton solutions}
\label{sec:model}
\bigskip

Using dimensionless variables and parameters, the Lagrangian of our model
reads (note that scalar fields in $D=1+1$ are dimensionless in natural units)
\cite{Halcrow:2023eph},
\begin{equation}
\mathcal{L}=\frac{m^2}{v^2}\left[\frac{1}{2}\partial_\mu \phi_i \partial^\mu\phi_i
-\left(1-\cos\phi_1\right)-\frac{\mu_1^2}{8}\left(1-\frac{2\phi_2}{\mu_2}
-\cos\frac{\phi_1}{2}\right)^2\right]\,,
\label{eq:lag}\end{equation}
for the scalar fields $\phi_1$ and $\phi_2$.
Here $m$ is the physical mass of $\phi_1$ such that the associated
dimensionless mass is one. Furthermore $v$ is a dimensionless model
parameter such that the allowed physical vacua are
$\langle\phi_1\rangle=2\pi m/v$ with integer~$m$. The overall factor
has no effect on the classical dynamics, but is a relative weight for
the one-loop quantum corrections because it enters the relation
between the canonical momenta and the field velocities in canonical
quantization.  After taking that factor into account, it suffices to
work with the expression in square brackets. The first potential term
is taken from the sine-Gordon model and the second describes the
interaction between the two scalar fields $\phi_1$ and $\phi_2$.
Hence the coupling constant $\mu_1$ determines the strength of that interaction
while $\mu_2$ sets the scale for $\phi_2$. 

Denoting time derivatives by dots and spatial derivatives by primes, the
field equations obtained from Eq.~(\ref{eq:lag}) are
\begin{align}
\ddot{\phi}_1-\phi_1^{\prime\prime}&=-\sin\phi_1-\frac{\mu_1^2}{8}\left(1-\frac{2\phi_2}{\mu_2}
-\cos\frac{\phi_1}{2}\right)\sin\frac{\phi_1}{2}\cr
\ddot{\phi}_2-\phi_2^{\prime\prime}&=\frac{\mu_1^2}{2\mu_2}\left(1-\frac{2\phi_2}{\mu_2}
-\cos\frac{\phi_1}{2}\right)\,.
\label{eq:EQM} \end{align}
Vacuum solutions, constant fields for which the right-hand-sides of Eq.~(\ref{eq:EQM}) 
vanish and that minimize the field potential, come in two types, both labeled by an 
integer $n$:
\begin{equation}
\begin{array}{llll}
{\rm A}:\qquad\qquad \phi_1&=(4n+2)\pi & \quad {\rm and}\qquad \phi_2=\mu_2
&\qquad {\rm with }\qquad n=0,1,\ldots \cr
{\rm B}:\qquad\qquad \phi_1&=4n\pi & \quad {\rm and}\qquad \phi_2=0
&\qquad {\rm with }\qquad n=1,2,\ldots\,. \cr
\end{array}
\label{eq:vacua}
\end{equation}
As in Ref. \cite{Halcrow:2023eph}, we take the boundary conditions such that both fields 
vanish at negative spatial infinity and assume any of the vacua from
Eq.~(\ref{eq:vacua}) at positive infinity.
We can then assign the topological charge
\begin{equation}
N=\frac{\phi_1(\infty)-\phi_1(0)}{2\pi}=
\begin{dcases} 2n+1=1,3,5,\ldots & \quad \mbox{for case A}\\[1mm]
2n=2,4,6,\ldots & \quad \mbox{for case B.}\end{dcases}
\label{eq:charge}\end{equation}
The static soliton equations are
\begin{align}
\phi^{\prime\prime}_1&=\sin\phi_1+\frac{\mu_1^2}{8}\left(1-\frac{2\phi_2}{\mu_2}
-\cos\frac{\phi_1}{2}\right)\sin\frac{\phi_1}{2}\cr
\phi^{\prime\prime}_2&=-\frac{\mu_1^2}{2\mu_2}\left(1-\frac{2\phi_2}{\mu_2}
-\cos\frac{\phi_1}{2}\right)\,.
\label{eq:static} \end{align}

When the soliton approaches a vacuum of type A, the parameterization
$\phi_1=\overline{\phi}_1+(2n+1)\pi$ and $\phi_2=\overline{\phi}_2+\frac{\mu_2}{2}$
shows that both $\overline{\phi}_i$ are odd functions of the coordinate. Then the 
initial conditions at $x\sim0$ are
\begin{equation}
\phi_1(x)\sim(2n+1)\pi+ax\qquad {\rm and}\qquad \phi_2(x)\sim \frac{\mu_2}{2}+bx\,.
\label{eq:initialB}\end{equation}
The asymptotic behavior as $x\to \infty$ for the type A boundary conditions is
\begin{equation}
\phi_1(x)\sim (4n+2)\pi+A{\rm e}^{-x}
\qquad {\rm and}\qquad 
\phi_2(x)\sim \mu_2+ B{\rm e}^{-\mu_0 x}
\quad{\rm where}\quad \mu_0={\rm min}\left(2,\frac{\mu_1}{\mu_2}\right)\,.
\label{eq:asympa}\end{equation}
The parameters $a$, $b$, $A$ and $B$ will be tuned in a shooting
method such that $\phi_i$ and $\phi^\prime_i$ are continuous
functions. That is, for an initial guess of these parameters we
numerically integrate Eq.~(\ref{eq:static}) both from $x=0$ to an 
intermediate matching point $x_{\rm m}\gg0$ and from $x_{\rm max}\gg
x_{\rm m}$ to $x_{\rm m}$. We then apply a Newton algorithm for the
parameters to obtain  continuous functions at $x_{\rm m}$.

When the soliton interpolates between two vacua of type B, the parameterizations
$\phi_1=\overline{\phi}_1+2n\pi$ and $\phi_2=\overline{\phi}_2$
are consistent with taking $\overline{\phi}_1$
and $\overline{\phi}_2$ to be odd and even functions, respectively.
Hence we can solve the static equations on the half-line $x\ge0$ with
the initial conditions around $x\sim0$
\begin{equation}
\phi_1(x)\sim 2n\pi+ax\qquad {\rm and}\qquad \phi_2(x)\sim b+\mathcal{O}(x^2)\,.
\label{eq:initialA}\end{equation}
In this case the large $x$ asymptotic behavior is found to be
\begin{equation}
\phi_1(x)\sim 4n\pi+A{\rm e}^{-x}
\qquad {\rm and}\qquad 
\phi_2(x)\sim B{\rm e}^{-\mu_0 x}\,.
\label{eq:asympb}\end{equation}
Again  $a$, $b$, $A$ and $B$ will be tuned to construct the soliton.

\begin{figure}
\centerline{\epsfig{file=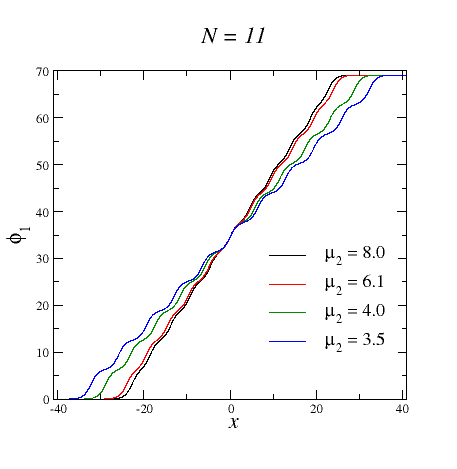,width=7cm,height=5cm}\hspace{1cm}
\epsfig{file=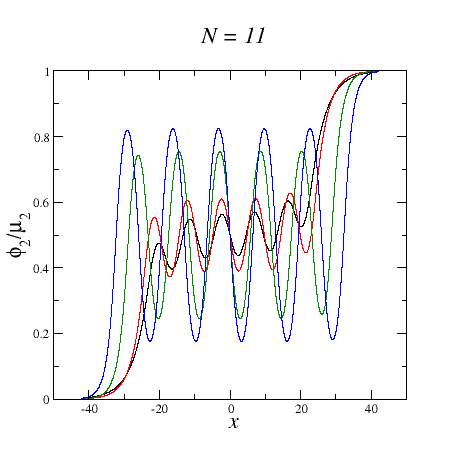,width=7cm,height=5cm}}
\caption{\label{fig:N11a}Profiles for $N=11$: left panel $\phi_1$; 
right panel $\phi_2$ normalized to $\mu_2$. Here $\mu_1=2.0$}
\end{figure}

\begin{figure}
\centerline{\epsfig{file=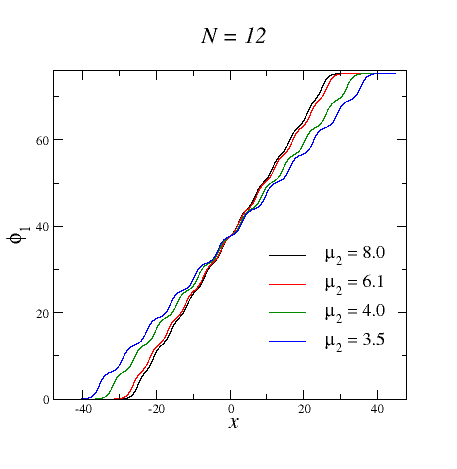,width=7cm,height=5cm}\hspace{1cm}
\epsfig{file=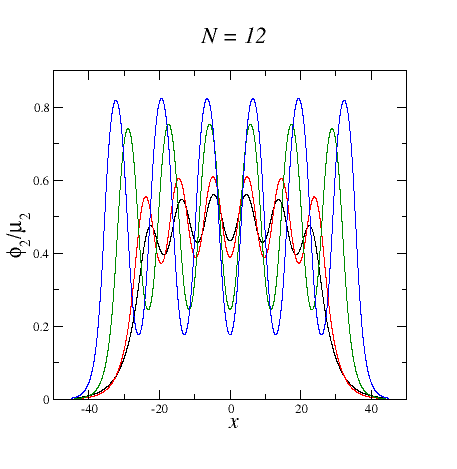,width=7cm,height=5cm}}
\caption{\label{fig:N12a}Profiles for $N=12$: $\phi_1$ (left panel)
and $\phi_2$ normalized to $\mu_2$ (right panel). Here $\mu_1=2.0$.}
\end{figure}

\begin{figure}
\centerline{\epsfig{file=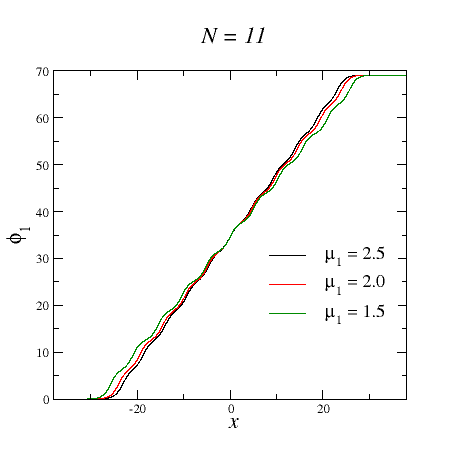,width=7cm,height=5cm}\hspace{1cm}
\epsfig{file=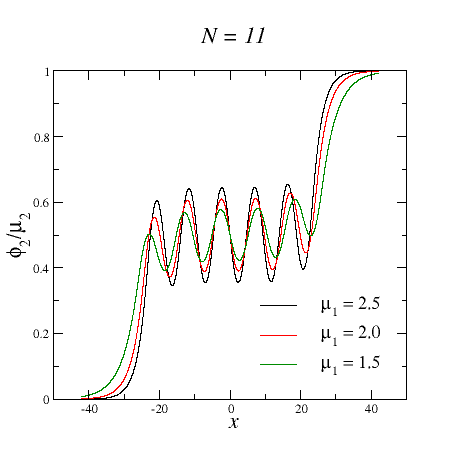,width=7cm,height=5cm}}
\caption{\label{fig:N11b}Profiles for $N=11$: $\phi_1$ (left panel)
and $\phi_2$ normalized to $\mu_2$ (right panel). Here $\mu_2=6.1$.}
\end{figure}

\begin{figure}
\centerline{\epsfig{file=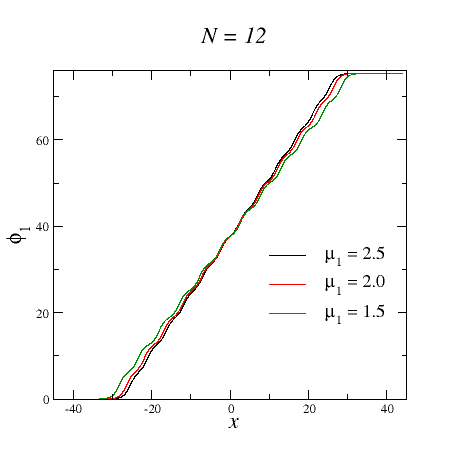,width=7cm,height=5cm}\hspace{1cm}
\epsfig{file=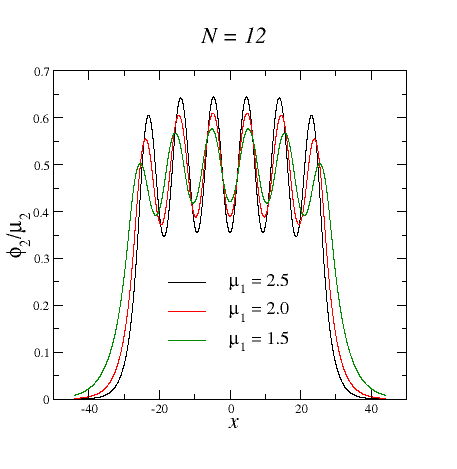,width=7cm,height=5cm}}
\caption{\label{fig:N12b}Profiles for $N=12$: $\phi_1$ (left panel)
and $\phi_2$ normalized to $\mu_2$ (right panel). Here $\mu_2=6.1$.}
\end{figure}

Finding solutions to Eqs.~(\ref{eq:static}) requires delicate choices of
the numerical parameters (mainly $x_{\rm m}$ and $x_{\rm max}$).
So it is advisable to store  them together with the
parameters from the shooting method, especially
for higher topological charges; those configurations are quite wide, and
one has to find the coefficients $A$ and $B$ in Eqs.~(\ref{eq:initialB})
and~(\ref{eq:initialA}) to high precision, because
the exponential is tiny for very large $x$ but
its product with these coefficients must be of $\mathcal{O}(1)$.
Numerical solutions for the profile functions have already been presented
in Ref.~\cite{Halcrow:2023eph}. Here we will focus on the structure
of solutions for higher topological charges, as shown in Figures
\ref{fig:N11a} and \ref{fig:N12a} for different values of the model 
parameter $\mu_2$ and in Figures \ref{fig:N11b} and \ref{fig:N12b}
of various values of $\mu_1$.

For larger topological charges, the profile function $\phi_2$ 
shows that the solitons resemble distinct unit charge sine-Gordon solitons 
when $\mu_2$ is small, but they turn into single lumps in which
the solitons sit on top of each other when $\mu_2$ is large.
On the other hand, the profiles show only little variation with $\mu_1$.

Numerical results for the classical energy 
\begin{equation}
E_{\rm cl}^{(N)}=\int_{0}^\infty dx\left[\phi_1^{\prime2}+\phi_2^{\prime2}
+2\left(1-\cos\phi_1\right)+\frac{\mu_1^2}{4}\left(1-\frac{2\phi_2}{\mu_2}
-\cos\frac{\phi_1}{2}\right)^2\right]
\label{eq:Ecl}\end{equation}
are listed in Table \ref{tab:ecl} for a set of model parameters, a
subset of which was also considered in Ref.~\cite{Halcrow:2023eph}. We
agree with their results and their numerical fit for $(E^{(N)}_{\rm
  cl}-NE^{(1)}_{\rm cl})/NE^{(1)}_{\rm cl}$,
where $E^{(1)}_{\rm cl}$ is the energy of a single soliton. For
comparison we note that the classical energy of the sine-Gordon
soliton is $E^{\rm sG}=8$ for the present units.  Derrick's
theorem~\cite{Derrick:1964ww} implies that the derivative and
non-derivative components should contribute equally in the above
integral. This has been used as the main criterion to validate the solutions.

The main effect of the $\phi_2$ field is to stabilize the multi-sine-Gordon
profile $\phi_1$. Without that second field, the multi-sine-Gordon configurations 
would split up into single solitons with infinite separation. This effect can 
also seen from the classical energy: Taking the $\phi_1$ profile  for $N=4$ 
and setting $\mu_1=0$ and $\phi_2^\prime=0$ yields an energy of $57.7$, while 
the exact classical energy is only moderately larger at $66.7$.

\begin{table}
\begin{tabular}{c|cccccccccccc}
$\mu_2$ & 1 & 2 & 3 & 4 & 5 & 6 & 7 & 8 & 9 & 10 & 11 & 12 \\
\hline
$4.0$ & 9.28 & 18.42 &  27.57 &  36.72 &  45.87 &  55.03 
& 64.18 & 73.33 & 82.48 & 91.63 & 100.78 & 109.93\cr
$6.1$ & 10.24 & 19.44 & 28.95 & 38.37 & 47.82 & 57.26 
& 66.71 & 76.14 & 85.59 & 95.03 & 104.47 & 113.92\cr
$8.0$ & 11.14 &  19.99 &  29.81 &  39.24 &  48.83 &  58.35 
&  67.90 &  77.44 &  86.99 &  96.53 &  106.07 &  115.62
\end{tabular}
\caption{\label{tab:ecl}The classical energy as a function of 
the topological charge $N$ (top line) for $\mu_1=2$.}
\end{table}

\begin{figure}
\centerline{\epsfig{file=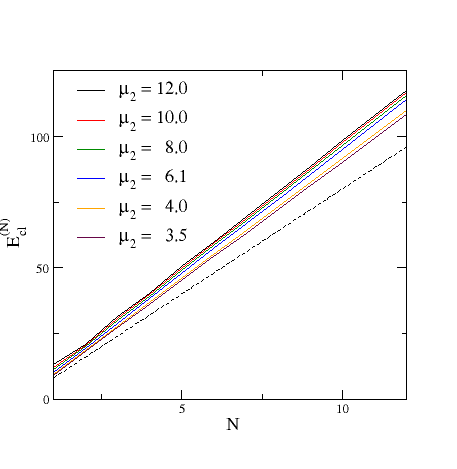,width=7cm,height=5cm}\hspace{1cm}
\epsfig{file=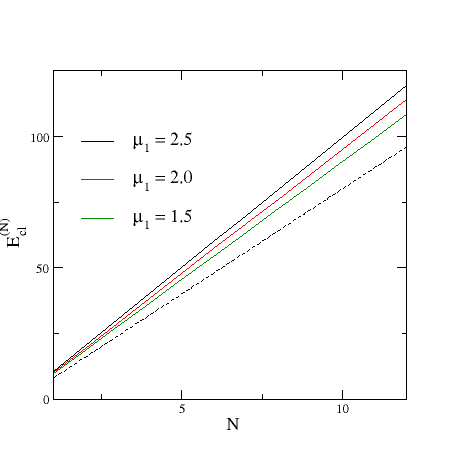,width=7cm,height=5cm}} 
\caption{\label{fig:Ecl}The classical energy, Eq.~(\ref{eq:Ecl}), for
$\mu_1=2.0$
and several values of $\mu_2$ (left panel) and for  $\mu_2=6.1$ and several
values of $\mu_1$ (right panel). The dashed line shows the classical energy, 
$8N$, of $N$ isolated sine-Gordon solitons.}
\end{figure}

\begin{figure}
\centerline{\epsfig{file=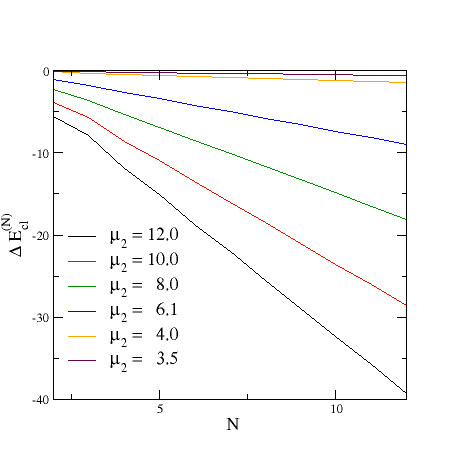,width=7cm,height=5cm}\hspace{1cm}
\epsfig{file=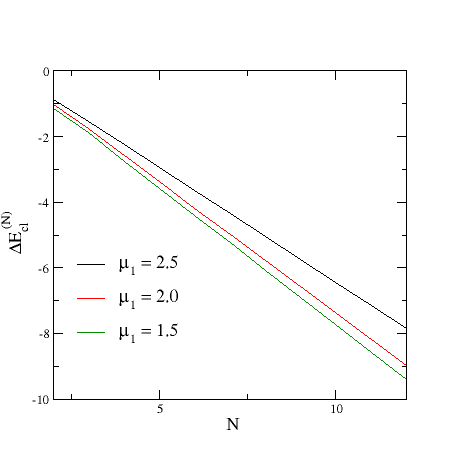,width=7cm,height=5cm}}
\caption{\label{fig:eclbind}The classical binding energy,
Eq.~(\ref{eq:eclbind}), for $\mu_1=2.0$
and several values of $\mu_2$ (left panel) and for  $\mu_2=6.1$ and several
values of $\mu_1$ (right panel).}
\end{figure}

It is more interesting to view the classical energies as functions
of the topological charge as shown in Figure \ref{fig:Ecl}.
Essentially we find a linear dependence with a small offset and only
small deviations, for low topological charges
($N\le3$) and for scenarios when the solitons form lumps. As we move
towards isolated solitons, we approach the limit $E^{(N)}_{\rm cl}\,\to\,8N$. 
The corresponding (negative) binding energies
\begin{equation}
\Delta E _{\rm cl}^{(N)} =E_{\rm cl}^{(N)}-NE_{\rm cl}^{(1)}
\label{eq:eclbind}\end{equation}
are shown in Figure \ref{fig:eclbind}. For the sets of model
parameters considered we find that $\Delta E _{\rm cl}^{(N)}$
decreases monotonically with the  topological charge, so that the
higher charge solutions are stable against decay into isolated
solitons with the same total charge. Furthermore we
observe that the (negative) gradient of this decrease is larger for cases
that have more compact soliton profiles.

We conclude this section by mentioning that there is a significant variation of 
$E _{\rm cl}^{(N)}$ with $\mu_1$, even though the profiles are not too
sensitive to that parameter. This variation arises from the $\mu_1^2$
factor in Eq.~(\ref{eq:Ecl}). However, the (almost) linear dependence
on $N$ was also observed for all considered values of $\mu_1$.

\section{Wave equations}
\bigskip

We linearize the field equations, Eq.~(\ref{eq:EQM}), by the ansatz
\begin{equation}
\phi_i(x,t)=\phi_i^{(s)}(x)+\eta_i(x){\rm e}^{-\imu \omega t}
\qquad {\rm for}\quad i=1,2\,.
\label{eq:lin1}
\end{equation}
The superscripts refer to the soliton solutions constructed above and we have 
omitted the frequency argument for the small amplitude fluctuations
$\eta_i(x)$.  The wave equations for these fluctuations can be written as
\begin{equation}
\eta_i^{\prime\prime}=-\omega^2\eta_i+M_{ij}\eta_j+V_{ij}\eta_j\,
\label{eq:lin2}
\end{equation}
with the mass and potential matrices given by~\cite{Halcrow:2023eph}
{\small
\begin{equation}
M=\begin{pmatrix}1 & 0 \cr 0 & \frac{\mu_1^2}{\mu_2^2}\end{pmatrix}
\qquad {\rm and}\qquad
V=\begin{pmatrix}\cos\phi^{(s)}_1-1
-\frac{\mu_1^2}{\mu_2}\left[\mu_2\cos\phi^{(s)}_1
-(\mu_2-2\phi^{(s)}_2)\cos\frac{\phi^{(s)}_1}{2}\right]\quad
&-\frac{\mu_1^2}{4\mu_2} \sin\frac{\phi^{(s)}_1}{2} \cr
-\frac{\mu_1^2}{4\mu_2} \sin\frac{\phi^{(s)}_1}{2} & 0\end{pmatrix}\,.
\label{eq:lin3} \end{equation}}
There is no explicit dependence on the topological charge $N$ because
it is fully encoded in the potential $V$. Therefore we will not add
the label $N$ to the solutions of Eq.~(\ref{eq:lin2}), but that
dependence is understood in what follows.

The case when $\phi^{(s)}_1/2$ is an even multiple of $\pi$ and $\phi^{(s)}_2=0$ 
obviously has $V_{11}=0$. In the other case when $\phi^{(s)}_1/2$ is an
odd multiple of $\pi$ and $\phi^{(s)}_2=\mu_2$ yields $\mu_2+(\mu_2-2\mu_2)=0$ for 
the expression in square brackets contained in $V_{11}$. Hence we have equal 
masses not only at positive and negative spatial infinity but also for the two cases 
of even and odd topological charges, because in either scenario the potential matrix 
vanishes asymptotically.

For \underline{case A} (odd topological charges), spatial reflections impose
\begin{equation}
\phi^{(s)}_1(-x)=2(2n+1)\pi-\phi^{(s)}_1(x)
\qquad {\rm and}\qquad
\phi^{(s)}_2(-x)=\mu_2-\phi^{(s)}_2(x)\,.
\label{eq:refA1}
\end{equation}
With 
$$
(\mu_2-2\phi^{(s)}_2)\cos\frac{\phi^{(s)}_1}{2}\stackrel{x\to-x}{\longrightarrow}
(\mu_2-2\mu_2+2\phi^{(s)}_2)\cos\left[\pi-\frac{\phi^{(s)}_1}{2}\right]
=(-\mu_2+2\phi^{(s)}_2)(-1)\cos\frac{\phi^{(s)}_1}{2}
$$
it is obvious that $V_{11}(-x)=V_{11}(x)$. Furthermore
$$
\sin\frac{\phi^{(s)}_1}{2}\stackrel{x\to-x}{\longrightarrow}
\sin\left[\pi-\frac{\phi^{(s)}_1}{2}\right]
=\sin\frac{\phi^{(s)}_1}{2}
$$
is also symmetric. Hence soliton configurations with odd topological charges
induce a symmetric scattering problem. That is, for positive parity
both $\eta_1$ and $\eta_2$ are symmetric functions of the coordinate,
while for negative parity both fluctuations are anti-symmetric functions.
For \underline{case B} (even topological charges), spatial reflections impose
\begin{equation}
\phi^{(s)}_1(-x)=4n\pi-\phi^{(s)}_1(x)
\qquad {\rm and}\qquad
\phi^{(s)}_2(-x)=\phi^{(s)}_2(x)\,,
\label{eq:refB1}
\end{equation}
which straightforwardly shows that again $V_{11}(-x)=V_{11}(x)$. On the other hand
$$
\sin\frac{\phi^{(s)}_1}{2}\stackrel{x\to-x}{\longrightarrow}
\sin\left[2n\pi-\frac{\phi^{(s)}_1}{2}\right]
=-\sin\frac{\phi^{(s)}_1}{2}
$$
leads to $V_{12}(-x)=-V_{12}(x)$, {\it i.e.\@} even topological charges
come with a skew-symmetric scattering problem. Here $\eta_1$ and $\eta_2$
are respectively symmetric and anti-symmetric functions for positive
parity, and vice versa for negative parity.

We apply an adaptive step size control when numerically integrating Eq.~(\ref{eq:lin2}),
both for the bound and scattering states. Since the soliton is only known at 
discrete values of the coordinate, this requires a numerically costly
interpolation algorithm.

\section{Bound States}

Although we do not require the bound states explicitly for our
calculation, they can be of use in analyzing properties of the
classical background, particularly in distinguishing between isolated
solitons and combined lumps. To construct the bound state wavefunctions it 
is again sufficient to only consider  the half-line $x\ge0$. The boundary 
conditions at $x\,\to\,\infty$ are the same for all parities and topological
charges. In that limit bound states require $\eta^\prime_i=-\kappa_i\eta_i$, 
where $\kappa_i$ is the (imaginary) wavenumber associated with the bound
state energy $\omega_b=\sqrt{1-\kappa_1^2}=\sqrt{\mu^2-\kappa_2^2}$. We 
only consider cases with $\mu_1<\mu_2$ and
therefore the threshold mass is given by $\mu=\frac{\mu_1}{\mu_2}$. Since
the fluctuation equations are linear and the potential matrix vanishes
at spatial infinity, we may write the linearly independent solutions
for sufficiently large $x_{\rm max}$ as
\begin{align}
\eta^{(1)}_1(x_{\rm max})&=1\,,\quad \eta^{(1)\prime}_1(x_{\rm max})
=-\sqrt{1-\omega_b^2}
\quad {\rm and}\quad \eta_2^{(1)}(x_{\rm max})\equiv0 \cr
\eta^{(2)}_2(x_{\rm max})&=1\,,\quad \eta^{(2)\prime}_2(x_{\rm max})
=-\sqrt{\mu^2-\omega_b^2}
\quad {\rm and}\quad \eta_1^{(2)}(x_{\rm max})\equiv0\,.
\label{eq:bcr}\end{align}

However, for the boundary conditions at $x=0$ we have to distinguish 
between a number of cases. For odd $N$ the positive parity channel has
\begin{equation}
\eta^{(3)}_1(0)=1 \quad {\rm and}\quad \eta^{(4)}_2(0)=1\,,
\label{eq:bcl1}\end{equation}
while the negative parity channel has
\begin{equation}
\eta^{(3)\prime}_1(0)=1 \quad {\rm and}\quad \eta^{(4)\prime}_2(0)=1\,.
\label{eq:bcl2}\end{equation}
For even $N$ we impose mixed conditions. They read
\begin{equation}
\eta^{(3)}_1(0)=1 \quad {\rm and}\quad \eta^{(4)\prime}_2(0)=1\,,
\label{eq:bcl3}\end{equation}
for the positive parity channel
and 
\begin{equation}
\eta^{(3)\prime}_1(0)=1 \quad {\rm and}\quad \eta^{(4)}_2(0)=1\,,
\label{eq:bcl4}\end{equation}
for the negative parity channel.
Functions and derivatives that are not explicitly listed in 
Eqs.~(\ref{eq:bcl1})-(\ref{eq:bcl4}) are set to zero at $x=0$. The above boundary 
conditions ensure that the wavefunctions are regular when integrating to a 
common matching point $x_{\rm m}\in\left[0,x_{\rm max}\right]$. We then have to 
construct linear combinations such that the wavefunctions and their first derivatives
are continuous at $x_{\rm m}$. Such continuous combinations exist when the 
determinant of
\begin{equation}
D(\omega)=\begin{pmatrix}
\eta_1^{(1)} & \eta_1^{(2)} & \eta_1^{(3)} & \eta_1^{(4)}\\[0mm]
\eta_2^{(1)} & \eta_2^{(2)} & \eta_2^{(3)} & \eta_2^{(4)}\\[0mm]
\eta_1^{(1)^\prime} & \eta_1^{(2)^\prime} & \eta_1^{(3)^\prime} & \eta_1^{(4)^\prime}\\[0mm]
\eta_2^{(1)^\prime} & \eta_2^{(2)^\prime} & \eta_2^{(3)^\prime} & \eta_2^{(4)^\prime}
\end{pmatrix}_{x=x_{\rm m}}
\label{eq:match}\end{equation}
vanishes. We have made explicit the dependence on the single particle
energy $\omega$ that arises from solving Eq.~(\ref{eq:lin2}) with the
boundary condition, Eq.~(\ref{eq:bcr}). We then scan this determinant
for $\omega\in\left[0,\mu\right]$ and identify the bound state
energies from ${\rm det}\left[D(\omega_j)\right]=0$. The numerical
results are corroborated by varying $x_{\rm m}$ and $x_{\rm max}$ in
appropriate intervals.

The standard sine-Gordon soliton only has a single bound state, the translational zero mode.
Hence for cases with wide solitons ($\mu_2$ small) we expect $N$ bound states (one of which 
is again the translational zero mode\footnote{Numerical error often causes 
that zero mode to be found at $\omega_0\approx 10^{-3}\imu$.}) near zero
energy. 
As $\mu_2$ increases, these energy eigenvalues should increase as well. A bit surprisingly, 
the bound states do not necessarily alternate between positive and negative channels. However,
the number of bound states in the latter never exceeds that in the former (including the 
zero mode in the counting). 

\begin{table}
\begin{tabular}{cc|cc|cc|cc}
\multicolumn{2}{c|}{$\mu_2=12.0$}  & \multicolumn{2}{c|}{$\mu_2=10.0$} 
& \multicolumn{2}{c|}{$\mu_2=8.0$} & \multicolumn{2}{c}{$\mu_2=3.5$} \cr
\hline
$\omega$ & $\omega/\mu$ & $\omega$ & $\omega/\mu$ 
& $\omega$ & $\omega/\mu$ & $\omega$ & $\omega/\mu$\cr
\hline
0.102 & 0.613 & 0.100 & 0.501 & 0.095 & 0.378 & 0.030 & 0.053\cr
0.129 & 0.773 & 0.148 & 0.739 & 0.171 & 0.684 & 0.058 & 0.101\cr
 --   &  --   & 0.190 & 0.950 & 0.182 & 0.728 & 0.080 & 0.140\cr
 --   &  --   & 0.194 & 0.972 & 0.204 & 0.816 & 0.094 & 0.165\cr
 --   &  --   &  --   &  --   & 0.235 & 0.938 & 0.101 & 0.177\cr
\hline 
0.052 & 0.315 & 0.051 & 0.255 & 0.048 & 0.192 & 0.015 & 0.027\cr
0.149 & 0.894 & 0.147 & 0.737 & 0.139 & 0.557 & 0.045 & 0.079\cr
0.151 & 0.904 & 0.166 & 0.832 & 0.185 & 0.740 & 0.070 & 0.122\cr
 --   &  --   &  --   &  --   & 0.220 & 0.880 & 0.088 & 0.154\cr
 --   &  --   &  --   &  --   & 0.224 & 0.895 & 0.098 & 0.172\cr
\end{tabular}
\caption{\label{tab:bsN11}Bound state energies for $N=11$ and $\mu_1=2.0$. The top and bottom brackets 
are positive and negative parity, respectively. The zero mode in the positive parity channel is not
explicitly listed.}
\end{table}

\begin{table}
\begin{tabular}{cc|cc|cc|cc}
\multicolumn{2}{c|}{$\mu_2=12.0$}  & \multicolumn{2}{c|}{$\mu_2=10.0$} 
& \multicolumn{2}{c|}{$\mu_2=8.0$} & \multicolumn{2}{c}{$\mu_2=3.5$} \cr
\hline
$\omega$ & $\omega/\mu$ & $\omega$ & $\omega/\mu$ 
& $\omega$ & $\omega/\mu$ & $\omega$ & $\omega/\mu$\cr
\hline
0.094 & 0.565 & 0.092 & 0.461 & 0.087 & 0.348 & 0.028 & 0.049\cr
0.147 & 0.880 & 0.163 & 0.817 & 0.168 & 0.672 & 0.054 & 0.094\cr
 --   &  --   & 0.179 & 0.893 & 0.183 & 0.730 & 0.075 & 0.131\cr
 --   &  --   &  --   &  --   & 0.218 & 0.873 & 0.090 & 0.158\cr
 --   &  --   &  --   &  --   & 0.232 & 0.926 & 0.099 & 0.173\cr
\hline
0.048 & 0.289 & 0.047 & 0.234 & 0.044 & 0.176 & 0.014 & 0.024\cr
0.128 & 0.768 & 0.136 & 0.680 & 0.128 & 0.514 & 0.041 & 0.072\cr
0.138 & 0.830 & 0.147 & 0.735 & 0.170 & 0.682 & 0.065 & 0.114\cr
 --   &  --   & 0.186 & 0.930 & 0.200 & 0.799 & 0.084 & 0.146\cr
 --   &  --   &  --   &  --   & 0.205 & 0.820 & 0.095 & 0.167\cr
 --   &  --   &  --   &  --   & 0.232 & 0.930 & 0.101 & 0.177
\end{tabular}
\caption{\label{tab:bsN12}Same as Tab. \ref{tab:bsN11} for $N=12$.}
\end{table}

In Tables \ref{tab:bsN11} and \ref{tab:bsN12} we present some representative data for
the bound state energies. Indeed their absolute values increase as the soliton becomes
more compact. However, the main effect is the decrease of the threshold energy $\mu$
so that bound states disappear into the continuum. The structure of the bound state 
spectrum significantly changes with $\mu_2$. Eventually, when $\mu_2$ is tuned such that
the large $N$ soliton is merely a lump, only the zero mode in the positive parity 
channel will remain bound.

\section{Vacuum polarization energy}

We now turn to our main objective: calculating the leading quantum
corrections to these solitons' classical energies. This is the vacuum
polarization energy (VPE), computed as the renormalized sum of the
shift in the zero-point energies
\begin{equation}
E_{\rm VPE}=\frac{1}{2}\sum_k \left[\omega_k-\omega_k^{(0)}\right]+E_{\rm CT}\,.
\label{eq:defVPE}\end{equation}
The $\omega_k$ are the energy eigenvalues in Eq.~(\ref{eq:lin2}) and 
$\omega_k^{(0)}$ are their counterparts for $V\equiv0$. The counterterm 
contribution, $E_{\rm CT}$ implements the renormalization. We follow the
spectral approach~\cite{Graham:2009zz} which is based on two
important features. First, the continuum part of the sum in
Eq.~(\ref{eq:defVPE}) is expressed as momentum integral weighted by
the change in the density of states induced by the soliton. This
weight is written in terms of the scattering data associated with
Eq.~(\ref{eq:lin2}). Second, these scattering data are expanded in
powers of $V$ and the leading terms of that expansion are subtracted
under the momentum integral to render it finite. These subtractions
are added back in the form of Feynman diagrams, which are combined
with $E_{\rm CT}$ to yield finite expressions.
This approach is particularly efficient when continuing the scattering
problem  for the momentum $k$ as defined by the dispersion relation
$\omega=\sqrt{k^2+\mu^2}$ to the  imaginary axis $k=\imu t$ with
real~$t\in\left[\mu,\infty\right]$, in which case the bound states no
longer enter explicitly. Ref.~\cite{Graham:2022rqk} gives a
recent review of spectral methods while a number of obstacles for the
real momentum  formulation has been recently analyzed in
Ref.~\cite{Petersen:2024krb}.

Here we will refer to and follow Ref.~\cite{Weigel:2017kgy} 
for the application of the spectral method to a two component theory
with different masses.  In the notation of that reference, odd
topological charges give rise to a symmetric scattering problem while
that for even charges is skew-symmetric. The central component of the
calculation is the computation of the Jost matrix from the
differential equation
\begin{equation}
Z^{\prime\prime}(t,x)=2Z^\prime(t,x)D(t)+\left[M^2,Z(t,x)\right]+V(x)Z(t,x)
\qquad {\rm with}\qquad
D(t)=\begin{pmatrix} \widetilde{t} & 0 \cr 0 & t\end{pmatrix}
\label{eq:master}\end{equation}
and $\widetilde{t}=\sqrt{t^2-\mu^2+1}$, which is the (analytically continued) momentum of the heavier field.
The other matrices are defined in the context of Eq.~(\ref{eq:lin2}). This differential
equation arises from applying the linear differential operator from Eq.~(\ref{eq:lin2}) 
to the matrix 
$$
\eta(x)=
Z(t,x)\begin{pmatrix}{\rm e}^{-\widetilde{t}x} & 0 \cr 0 & {\rm e}^{-tx}\end{pmatrix}\,.
$$
The elements within a given column represent the two fields while the two columns refer 
to the possible scattering channels. Eq.~(\ref{eq:master}) is solved subject to the 
boundary condition $\lim_{x\to\infty}Z(t,x)=\ID$ and subsequently the Jost matrices
\begin{equation}
F_S(t)=\lim_{x\to0}\left[Z(t,x)-Z^\prime(t,x)D^{-1}(t)\right]
\qquad {\rm and}\qquad
F_A(t)=\lim_{x\to0}Z(t,x)
\label{eq:defjostsym}
\end{equation}
are extracted. For the symmetric scattering problem (case A) we next compute the
Jost function
\begin{equation}
\nu_A(t)={\rm ln}\,{\rm det}\left[F_S(t)F_A(t)\right]
\label{eq:JostA}\end{equation}
and find the vacuum polarization energy for odd $N$
\begin{equation}
E^{(N)}_{\rm VPE} \equiv 
\int_{\mu}^{\infty} \frac{dt}{2\pi} 
\frac{t}{\sqrt{t^2-\mu^2}}\,\left[\nu_A(t)-\nu^{(1)}(t)\right]+E_{\rm FD}+E_{\rm CT}
=\int_{\mu}^{\infty} \frac{dt}{2\pi} 
\frac{t}{\sqrt{t^2-\mu^2}}\,\left[\nu_A(t)-\nu^{(1)}(t)\right]\,.
\label{eq:Evac} \end{equation}
We have rendered the (imaginary) momentum integral finite by subtracting the
Born approximation (recall that $V_{22}(x)=0$ in this model)
\begin{equation}
\nu^{(1)}(t)=\frac{1}{\sqrt{t^2-\mu^2+1}}\int_{0}^\infty dx\, V_{11}(x)
\label{eq:Born1}\end{equation}
and adding it back as a Feynman diagram. In the no-tadpole scheme, that
full diagram (not just its ultraviolet divergent part) is canceled by
the counterterm. For case B with skew parity, we replace the Jost
matrices above by~\cite{Weigel:2017kgy}
\begin{equation}
F_{\pm}(t)=\left[P_{\pm}F_S(t)D_{\mp}(t)+P_{\mp}F_A(t)D_{\pm}^{-1}(t)\right]\,,
\label{eq:skjost}\end{equation}
with projectors  
$P_{+}=\begin{pmatrix}1 & 0 \cr 0 & 0\end{pmatrix}$ and
$P_{-}=\begin{pmatrix}0 & 0 \cr 0 & 1\end{pmatrix}$
and factor matrices
$D_{+}(t)=\begin{pmatrix}-\widetilde{t} & 0 \cr 0 & 1\end{pmatrix}$ and  
$D_{-}(t)=\begin{pmatrix}1 & 0 \cr 0 & -t\end{pmatrix}$.
From these expressions we finally compute the appropriate Jost function for  
imaginary momenta
\begin{equation}
\nu_B(t) \equiv {\rm ln}\,{\rm det}\left[F_{+}(t) \,F_{-}(t)\right]\,,
\label{eq:defnu} \end{equation}
which then replaces $\nu_A(t)$ in Eq.~(\ref{eq:Evac}) for even
$N$. The first-order Born approximation is not modified since it does
not involve off-diagonal elements of the potential matrix.

In Table \ref{tab:vpe} we present numerical results for the VPE,
choosing parameters to include some of those considered in
Ref.~\cite{Halcrow:2023eph}. The arxiv version of that
reference gives VPEs for $N=1$ and $N=2$ and two values of $\mu_2$:
$-0.302$ ($\mu_2=4.0$) and $-0.297$ ($\mu_2=6.1$) for $N=1$. For the other case, $N=2$ those 
authors list $-0.547$ and $-0.486$. Certainly, within potential numerical errors these results 
are consistent with the corresponding data in Table \ref{tab:vpe}. This result also provides 
a check of the skew parity procedure, since taking $\nu_A(t)$ for case B incorrectly 
yields $-0.522$ and $-0.452$. The approach of Ref.~\cite{Halcrow:2023eph} is based on an 
expression from Ref.~\cite{Cahill:1976im}, which starts from Hermitian operators $H^2$ and $H_0^2$ 
with eigenvalues $\omega_k^2$ and $\omega^{(0)2}_k$, respectively. These operators are
related as  $H^2=H_0^2+V$ so that the VPE with the no-tadpole condition is\footnote{The
same expression for the VPE was obtained in Ref.~\cite{Evslin:2019xte} using a different and
quite intricate formalism. In Refs.~\cite{Halcrow:2023eph,Cahill:1976im,Evslin:2019xte} 
the no-tadpole scheme is referred to as normal-ordering.}
\begin{equation}
E_{CCG}=\frac{1}{2}\sum_k\left(\omega_k-\omega^{(0)}_k\right)\Big|_{\mbox{\tiny no tadp.}}
=\frac{1}{2}{\rm tr}\left[H-H_0-\frac{V}{2H_0}\right]
=-{\rm tr}\left[\frac{\left(H-H_0\right)^2}{4H_0}\right]\,.
\label{eq:CCG}\end{equation}
The evaluation of the trace requires the Fourier transforms of all eigenmodes of
the scattering problem, Eq.~(\ref{eq:lin2}) and a subsequent sum/integral over 
all modes. It is thus susceptible to numerical inaccuracies which may grow
as the soliton gets wider, making it more difficult to carry out the calculation 
for higher charges. Considering only $N=1$ and $N=2$ does not allow for a general 
statement on the charge dependence of the quantum corrections.

\begin{table}
\begin{tabular}{c|cccccccccccc}
$N$ & 1 & 2 & 3 & 4 & 5 & 6 & 7 & 8 & 9 & 10 & 11 & 12\\
\hline
$\mu_2=4.0$ & -0.303 & -0.549 & -0.805 & -1.061 & -1.317 & -1.574 & -1.830 & -2.084
& -2.343 & -2.600 & -2.857 & -3.113\\
$\mu_2=6.1$ & -0.297 & -0.487 & -0.713 & -0.930 & -1.152 & -1.371 & -1.594 & -1.810
& -2.036 & -2.257 & -2.478 & -2.700\\
$\mu_2=8.0$ & -0.295 & -0.459 & -0.684 & -0.882 & -1.095 & -1.301 & -1.511 & -1.719 
& -1.927 & -2.137 & -2.347 & -2.555
\end{tabular}
\caption{\label{tab:vpe}The vacuum polarization energy as a function of
the topological charge $N$ and the scale $\mu_2$. The other parameter is
always $\mu_1=2.0$. The cases $\mu_2=4.0$ and $\mu_2=6.1$ are also considered
in Ref.~\cite{Halcrow:2023eph} for $N=1$ and $N=2$ only.}
\end{table}

\begin{figure} 
\centerline{\epsfig{file=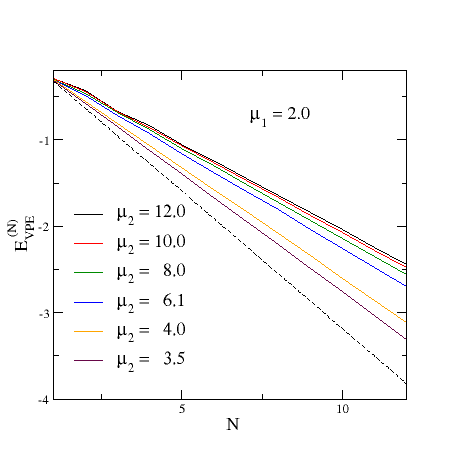,width=7cm,height=5cm}\hspace{1cm} 
\epsfig{file=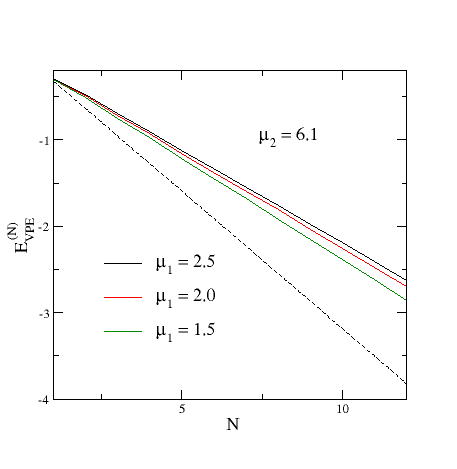,width=7cm,height=5cm}}
\caption{\label{fig:engs}Vacuum polarization energies for various model parameters.
The dashed lines refer to the case of $N$ isolated single sine-Gordon solitons.}
\end{figure}

Figure \ref{fig:engs} shows the resulting vacuum polarization energies as functions 
of the winding number $N$, Eq.~(\ref{eq:charge}).
Like the classical energies,
these are essentially straight lines, with small deviations for low $N$ and large $\mu_2$,
{\it i.e.\@} when the solitons sit on top of each other. This time, however, the straight
lines have a negative slope. More importantly, we do not see any correlation between
the bound state spectrum and this linear dependence. Regardless of whether the model parameters 
yield many strongly bound modes or only a few loosely bound ones, the functional dependence of
$E^{(N)}_{\rm VPE}$ is very similar. We consider this result to be a consequence of Levinson's
theorem~\cite{Barton:1984py}, which may have the interpretation that a state passing the 
threshold has compensating effects on the bound state and continuum contributions to the VPE. 
For comparison, we may consider the bound state contribution 
$E_{\rm b.s.}=\frac{1}{2}\sum_j^{\rm b.s.}\left(\omega_j-\mu\right)$ 
which was used to approximate the VPE in Ref. \cite{Gudnason:2023jpq}, though it
is explicitly included only for the real momentum formulation.
Taking, for example, $N=12$ and $\mu_1=2.0$, this quantity
changes drastically from $-3.055$ to $-0.223$ when going from $\mu_2=3.5$ to 
$\mu_2=12.0$. On the other hand the total VPE only changes from $-3.305$ to $-2.437$. In one case
the bound states give a reasonable approximation to the VPE, but in
the other their contribution is off by an 
order of magnitude. Even though $E_{\rm b.s.}$ is not exactly the bound state 
contribution in the trace of Eq.~(\ref{eq:CCG}), this short analysis
corroborates that the bound state spectrum alone is not necessarily a
reliable approximation to the VPE.

\section{Discussion}

As can be seen from Eq.~(\ref{eq:lag}), the Lagrange density scales with
$\frac{m^2}{v^2}$. Hence the classical energy scales like
$\frac{m}{v^2}$. On the other hand, the energy eigenvalues scale with
$m$ and so does the VPE. To one loop, the total energy in a given
topological sector is therefore proportional to
\begin{equation}
E_{\rm tot}^{(N)}=E^{(N)}_{\rm cl}+v^2E^{(N)}_{\rm VPE}\,.
\label{eq:etot}\end{equation}
For the one-loop approximation to be reliable, the magnitude of the first term should 
be substantially larger than that of the second. The numerical results from the previous
sections suggest that this inequality is still reasonably well fulfilled for values
as large as $v^2=\mathcal{O}(10)$. However, when we consider the (negative) binding energy
\begin{equation}
E_{\rm bind}^{(N)}=E_{\rm tot}^{(N)}-NE_{\rm tot}^{(1)}=
\left[E^{(N)}_{\rm cl}-NE^{(1)}_{\rm cl}\right]
+v^2\left[E^{(N)}_{\rm VPE}-NE^{(1)}_{\rm VPE}\right]\,.
\label{eq:ebind}\end{equation}
the classical and one-loop contributions may be of similar
magnitude.\footnote{In BPS models, the first term typically vanishes.}
It is therefore meaningful to study
$E_{\rm bind}^{(N)}$ as a function of the loop-counting parameter $v^2$.

In the previous sections we have seen that both $E^{(N)}_{\rm cl}$ and $E^{(N)}_{\rm VPE}$ 
are essentially linear functions of $N$. This implies that when $E_{\rm bind}^{(2)}<0$ we 
also have that $E_{\rm bind}^{(N)}<0$ for $N>2$ except for very specific values of $v^2$
that are sensitive to the small deviation from linearity.

\begin{figure}
\centerline{\epsfig{file=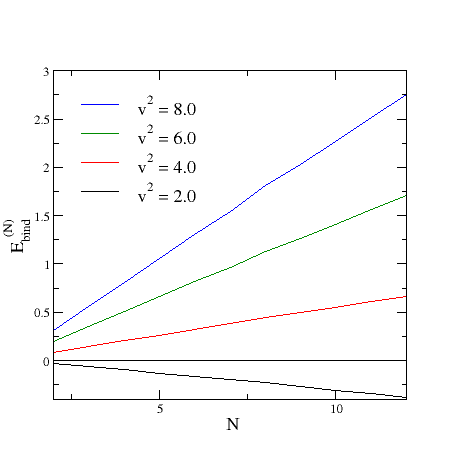,width=7cm,height=5cm}\hspace{1cm}
\epsfig{file=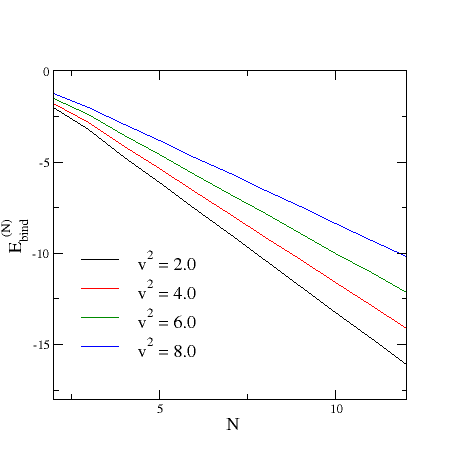,width=7cm,height=5cm}}
\caption{\label{fig:ebind}The one-loop binding energies, Eq.~(\ref{eq:ebind}) parameterized
by the loop-counting parameter $v^2$ for $\mu_1=2.0$ while $\mu_2=4.0$ (left panel) and 
$\mu_2=8.0$ (right panel). Note the different scale for the vertical axis in the two cases.}
\end{figure}

Fig.~\ref{fig:ebind} shows the total (negative) binding energies. The VPE
decreases the binding and eventually with a sufficiently large $v^2$
the multi-soliton configurations become unbound. 
For $\mu_2=4.0$ the limiting value is about $v_c^2=2.7$, while 
for $\mu_2=8.0$ it is substantially larger at $v_c^2=19$ ($\mu_1=2.0$ in both cases). 
As in the classical case, the more compact (larger $\mu_2$) 
the soliton, the stronger the binding. Indeed, these numerical
experiments suggest that large single mode binding energies 
are correlated with small soliton binding 
energies, because for larger $\mu_2$ and fixed $v^2$ the classical
energy becomes more dominant.

Due to the small offset in the linear behavior of the energies, for loop counting 
parameters in the close vicinity of $v_c^2$ it may occur that solitons with moderate 
topological charges are unstable while those with a larger one are
stable, but those binding energies are tiny.

Finally we compare this toy model with chiral soliton models
\cite{Weigel:2008zz}, such as the Skyrme model \cite{Skyrme:1961vq}, which treat baryons as solitons in an effective meson 
theory. Taking model parameters that are extracted from the meson properties, the lowest soliton 
mass exceeds the nucleon mass by about 30\% or more~\cite{Adkins:1983ya}.  It has been 
suggested that quantum corrections resolve that disagreement~\cite{Meier:1996ng}.
However, these models are not renormalizable (they even have quartic divergences) and so quantum 
corrections are ambiguous because assumptions about the un-matched counterterms
are needed.\footnote{In Ref. \cite{Walliser:1999ug} the winding number dependence of the VPE has been 
considered in the $D=2+1$, $O(3)$ version of the Skyrme model. Though the ultraviolet divergence
is only cubic, it is still not renormalizable.} Yet we can adopt that viewpoint in our model
and adjust $v^2$ to give a 20\% 
reduction to the ground state energy, {\it i.e.} $E_{\rm tot}^{(1)}\approx0.8E_{\rm cl}^{(1)}$, 
which would still be considered valid in the one-loop approximation. Taking again $\mu_1=2.0$, 
the resulting values are $v^2\approx 6.1$ for $\mu_2=4.0$ and $v^2\approx 7.6$ for $\mu_2=8.0$. 
Not surprisingly, the higher charge solitons become unbound in the former case but remain bound 
in the latter. Even though this consideration does not lead to a decisive conclusion, it shows 
that quantum corrections should be relevant for multi-baryon solitons~\cite{Feist:2012ps} in 
the Skyrme model. On the other hand, our results do not support the argument put forward 
in Ref.~\cite{Scholtz:1993jg} that quantum corrections could energetically favor $N$ isolated
unit charge solitons over a single soliton of charge $N$ when the latter has fewer than $N$ times 
as many zero modes (or strongly bound single particle modes) as the unit charge soliton.\footnote{That
conclusion emerged when truncating the trace in Eq.~(\ref{eq:CCG}) to its bound state component.
However, in $D=3+1$ the full trace is ultraviolet divergent \cite{Holzwarth:1994uj}.}
Our model finds such a bound state scenario for large $\mu_2$ when the compact soliton is not 
destabilized by quantum corrections due to the effects of the continuum.

\section{Conclusion}

We have considered an extension of the sine-Gordon model in $1+1$ dimensions
that has static soliton solutions with higher topological
charges. Depending on the model parameters, these solitons emerge
either as almost separated unit-charge configurations or as  compact
lumps with localized soliton profiles.

Our main objective has been to analyze the leading quantum corrections to the classical energies 
as a function of the topological charge. We have computed these corrections as the vacuum polarization 
energy (VPE), {\it i.e.} the renormalized sum of the energy shifts of the modes fluctuating about the 
soliton. This sum is ultraviolet divergent but in $D=1+1$ there is only a single divergent one-loop 
Feynman diagram. Within the no-tadpole renormalization scheme, that diagram is fully canceled by the 
counterterm. We have then applied spectral methods to compute the finite, non-perturbative VPE.
These methods make use of scattering data for fluctuations about the soliton, in particular the Jost 
function analytically continued to imaginary momenta. This formalism is extremely efficient and 
numerically not more laborious than constructing the soliton itself.

A main result of our numerical studies is that both the classical energy and its leading quantum
correction are essentially linear functions of the topological charge with a small offset. 
There are only minor deviations from linearity at small charges, even when the soliton appears 
as a single lump whose scattering properties are very different from those for a superposition of 
individual solitons. We have explicitly computed these energies for
topological charges less than
or equal to twelve. Since linearity is essentially exact for charges 
greater than or equal to six,
we conjecture that the straight line can safely be extended beyond twelve.

For the binding energies we find that the VPE may destabilize classically stable solitons 
when the single soliton structure is favored. To destabilize compact solitons,
the loop counting parameter must typically be taken so large that higher loop corrections 
cannot be omitted. However, for compact solitons the quantum
corrections considerably reduce the binding energies even for small or moderate values of the loop counting parameter.
We stress that it is the (small) offset that is relevant for binding,
since a purely linear dependence of the energy on the winding number
yields zero binding.

We have also explored the bound state spectrum of this scattering
problem. Generally soliton lumps have fewer bound states than isolated
solitons with the same total topological charge. Though the bound states energies
vary with the structure of the soliton, the main cause is the decrease of the
threshold as the soliton becomes more compact. The numerical results
suggest that this spectrum on its own provides little information
about the VPE, because it must be considered together with the
continuum contribution.

\acknowledgments
H.\ W.\ thanks F.\ G.\ Scholtz for clarifying comments on Ref.~\cite{Scholtz:1993jg}.
N.\ G.\ is supported in part by the
National Science Foundation (NSF) through grant PHY-2209582.
H.\ W.\ is supported in part by the National Research Foundation of
South Africa (NRF) by grant~150672.

\bibliographystyle{apsrev}

\end{document}